# Role of electrostatic potential energy in carbon nanotube augmented cement paste matrix

Muhammad Azeem[a] [*], Muhammad Azhar Saleem [b,c]

[a]*Department of APplied Physics and Astronomy, University of Sharjah, University City, Sharjah, 27272, United Arab Emirates.*
[b] *Department of Civil Engineering, University of Engineering and Technology, Lahore, Pakistan.*
[c] *American University of Sharjah, Sharjah, United Arab Emirates.*

## Abstract

The empirical data in conjunction with the quantum mechanical calculations show that the strength enhancement in the cement-carbon nanotubes (CNTs) composites is the courtesy of electrostatic potential energy. This is contrary to the general belief that the CNTs form bridges between the adjacent grains to slow down the breaking process. The yield point for the cement paste is improved up to 25% when prepared with 0.2 % by weight of various types of CNTs. A significant strength enhancement is observed with carboxyl functionalized (COOH) CNTs compared to other types. Further, an increase in the concentration of CNTs up to 0.4 wt% has a negative effect on the strength of the matrix. The electrostatic potential energy is mapped by using density functional theory (DFT) with ωB97X-D functional. At lower concentration of CNTs, ion-dipole interaction in the cement paste and the CNTs creates a very strong long range intermolecular force. Due to the increased entropy resulting from the exothermic hydration process, these forces augment the strength of the cement paste.

[*] Corresponding author. Tel.: +97165166764; fax: +0-000-000-0000 .
*E-mail address:* mazeem@sharjah.ac.ae

1. Introduction

The demand of cement is projected to rise at an unequivocal rate of 4.5% per year and has already reached 5.2 billion metric tons in 2019[1], making cement as one of the most consumable material by the humans. With the production of several billions of tons each year, the cement industry is one of the major contributor to the global $CO_2$ emissions[2]. However if we are able to substitute the part of the cement by a supplementary cementitious materials then its environmental impact can be controlled. A variety of the materials offer this possibility, nano-dimensional materials being one of them. The theoretically postulated fascinating quantum mechanical phenomena associated with the nano-structured materials are now being observed experimentally. Nanoparticles and nanotubes are highly active and have greater surface to volume ratio therefore more recently they have been pursued to give strength to the construction materials. The results are promising and show that the mixing of the nanomaterials in the cement paste makes it less prone to failure.

In general it is perceived that the nanomaterials occupy the pores in the hydrated cement paste and therefore act as binders between the grains[3,4]. There are, however, studies to suggest that the nanoparticles (particularly 3d metals) might occupy interstitial positions in the crystal structure of clinker[5,6] and thus accelerate the hydration process. Since most of the composites present in the hydrated cement are based on calcium and silicon, changing the ratio between two ions has significant effect on the physical properties of the cement composite[7]. The metallic nanoparticles may be used to replace the silicon atoms which then increases the ratio of calcium to silicon as well as changing their crystal structure. The same objective can also be achieved by using nanolimestone/$nanoCaCO_3$[6,8] and nanosilica[9]. There is also a speculation that the mixing of the carbon nanotubes and nanofibers may change the bonding order of the cement matrix[10]. An efficient acceleration in the hydration process of the cement paste is observed by the addition of nanoalumina and graphene oxide[6] and graphene[11,12] improving its compressive and flexural strength.

Carbon nanotube reinforced cementitious composites have attracted more attention because of their elastic modulus as high as 400 GPa [10] and tensile strength [0.4-5 GPa]. The Young's modulus for CNTs is around 1 TPa[13] making them superior in strength compared to steel. The fracture strain for CNTs is as high as 280, thanks to the carbon-carbon $sp^2$ bonding whereas fracture strain for steel is only around 20%. However notable is the fact that weak shear interaction between adjacent nanotubes may reduce their tensile strength. Therefore twisted CNT fibers in the form of strands experience significant reduction in their tensile strength. Further the defects introduced during the fabrication process also lead to reduced strength of CNTs. Aggregation of the CNTs permits ring opening mechanism and thus nucleation of the defects resulting in the concentration of the stress and strain at the site of defects. Therefore it is preferred that the CNTs are dispersed uniformly in the composite matrix allowing disentanglement of the tubes.

The mechanical properties of CNT doped cementitious materials depend on the several factors such as dispersion, proportion and length of CNTs. Various dispersion methods resulted in 14% to 71% increase in the strength of the cement mixtures[3,4]. Dispersion of longer CNTs is challenging therefore short length CNTs are preferred. Low concentration of well dispersed CNTs have produced effective enhancement of the mechanical properties of the cement.

Functional groups of CNTs (carboxyl –COOH and hydroxyl –OH) have also been found to affect the physical properties of the cement[4]. It is inferred from the experimental data that the hydrophilic carboxyl CNTs might initiate a chemical reaction between the carboxyl groups and the composites present in the cement. The cement paste with carboxyl functionalized tubes leads to lower concentration of tobermorite decreasing its strength considerably[14]. Porosity of the cement mixture is reduced with the increase in the CNTs concentration however, as discussed above and will be shown in the present work, an increase in the concentration does not necessarily mean an improved strength.

On the other hand, the mechanism of strength enhancement however is poorly emphasized and mostly inferences are made without any concrete proof. Crack bridging and the pore filling is the main reason identified for the cement matrix toughness. Therefore the present work highlights the underlying physics of the strengthening mechanism at the electronic level. It is shown that the modification of the cementitious properties strongly depends on the concentration of the foreign elements. It is, as a matter of fact, the electrostatic interaction that the augmented strength is obtained only for a certain concentration of the dopants. A change in the concentration of the added impurities affects the order of the electrostatic interaction with the hydrated cement paste therefore despite reduced porosity[10], the strength decreases with the increase in the concentration. The results are reported for the cement paste prepared with various concentrations (0.2 wt% and 0.4 wt%) of pristine, hydroxyl and carboxyl functionalized CNTs. Their compressive strength is tested after the gap of 7 and 28 days. The increase in the compressive strength is the highest for the samples with carboxyl functionalized nanotubes but the lowest for the hydroxyl functionalized. Further, a theoretical model is constructed to show that there is a strong electrostatic interaction between the CNTs and the $Ca(OH)_2$, the major component in the hydrated paste. It is illustrated that while the dipole moment of the CNTs is almost zero, the strain applied by the neighbouring crystallites distorts the geometry of CNTs and therefore introduce a change in the density of electronic charge on the surface of CNTs. A charge imbalance creates a strong dipole moment initiating the ion-CNTs dipole interaction. This is the origin of the electrostatic interaction between different composites in the cement paste and CNTs. The exothermic hydration process increases the entropy of the system and therefore the electrostatic interaction is long range, greatly improving the mechanical strength of the cement paste.

## 2. The Experiment

*2.1 Sample Preparation*

The cement used in preparation of the specimens was commonly available Type I ordinary Portland cement (OPC). The carbon nanotubes (CNTs) were obtained from Sisco Research Laboratories (SRL) and were of three types: pristine, hydroxyl functionalized (-OH) and carboxy (-COOH) functionalized. The average lengths of all three types of the CNTs was around 10-30 μm. The diameter of the pristine CNTs was around 8 nm whereas functionalized CNTs were of the diameter approximately 30-50 nm. The samples were tested for their tensile strength. Tap water was used for preparing the cement paste with a constant water-cement ratio of 0.4, for the entire stock of specimens. The equipment and the different stages of the sample preparation are exhibited in the Fig. 1(a-g).

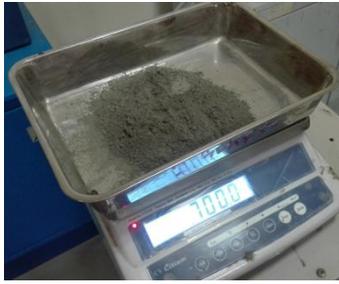 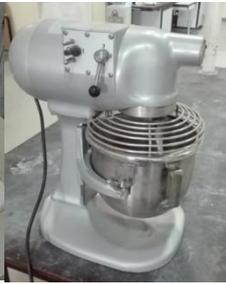 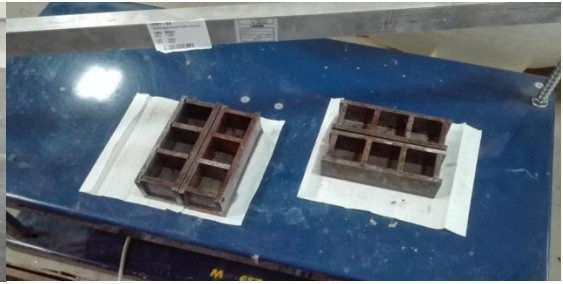

(a) Weighing of cement  (b) Bench mixer  (c) Empty Moulds Placed on Table Vibrator

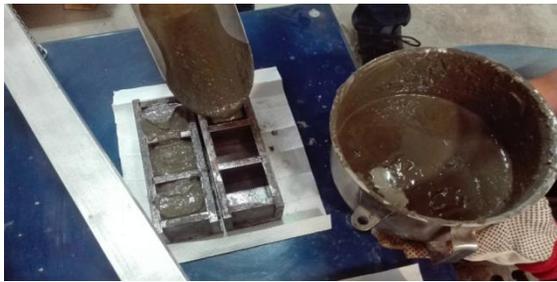 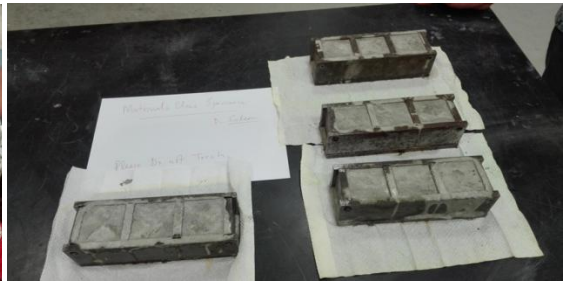

(d) Cement Paste being Poured in Moulds  (e) Specimens after 24 hrs

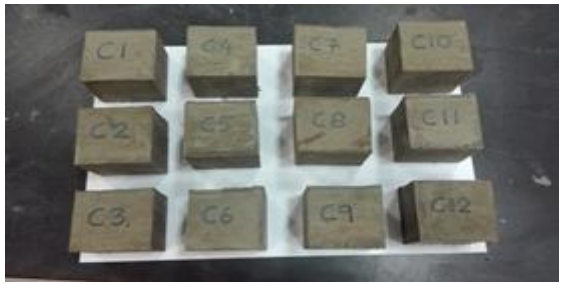 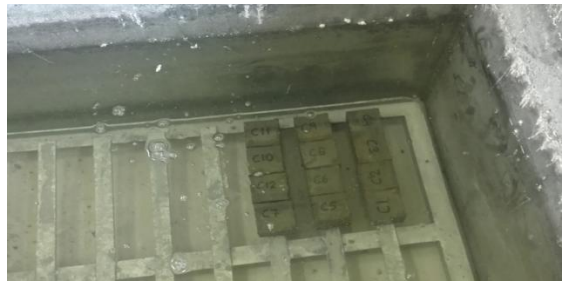

(f) Specimens after Demolding  (g) Specimens Placed in Water for Curing

Fig. 1 Various stages of sample preparation

Before using in the paste, cement was weighed, Fig. 1a, and sieved to get rid of any lumps. Mixing was carried out in a bench mixer, Fig. 1b. The CNTs and water were first added to the bowl and the mixer was switched on. Mixer was allowed to run for 2 mins to ensure that CNTs are uniformly dispersed in water. Afterwards, cement was added gradually while the mixing continued. Mixing was carried out for another 15 mins from the starting time of addition of cement. Gradual addition of cement avoided formation of cement lumps and resulted in consistent paste. After completion of mixing, cement paste was poured into the steel moulds of size 50 mm x 50 mm x 50mm, Fig. 1c & 1d, which were oiled to ensure easy removal. Table type external vibrator was used for compaction. Specimens were demolded after 24 hrs, Fig. 1(e, f), and then dipped in water until the day of testing, Fig.1 (g).

In total, 48 specimens were prepared and for each unique type three companions were cast to get an average compressive strength. Each specimen was a 50 mm cube, which is typically used to measure the compressive strength of cement. Four batches of cement paste were prepared. These batches comprised of the a batch of control specimens without CNTs, a batch each having 0.2% and 0.4% by dry weight of cement pristine CNTs, -OH and –COOH functionalized CNTs. The summary of the specimens prepared is presented in the Table 1. The specimens were labeled to indicate their type and composition. For example, letter C represents the control specimens, P-2 represents the specimens with pristine CNTs having 0.2 wt% and P-4 refers to the specimens with pristine CNTs having 0.4 wt%. Specimens with OH-CNT and COOH-CNTs were labeled accordingly.

**Table 1. Test Matrix**

| Type of Specimens | ID | No. of Specimens | Day of Testing | % Weight of CNT | Water/Cement Ratio |
|---|---|---|---|---|---|
| Control | C | 12 | 3, 7, 14, 28 | 0 | 0.4 |
| Pristine CNT | P-2 | 6 | 7, 28 | 0.2 | 0.4 |
| | P-4 | 6 | 7, 28 | 0.4 | 0.4 |
| OH-CNT | OH-2 | 6 | 7, 28 | 0.2 | 0.4 |
| | OH-4 | 6 | 7, 28 | 0.4 | 0.4 |
| COOH-CNT | COOH-2 | 6 | 7, 28 | 0.2 | 0.4 |
| | COOH-4 | 6 | 7, 28 | 0.4 | 0.4 |

Specimens were tested in load-control mode. Great care was taken to ensure that loading platens remain horizontal so that load gets uniformly distributed on the entire surface area of cubes. Testing of the specimens is exhibited in the Fig. 2(a, b).

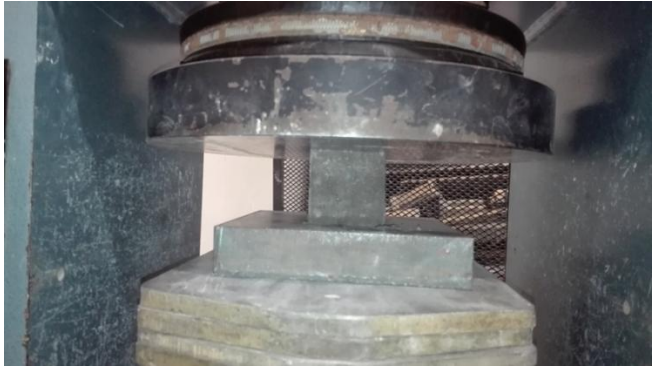
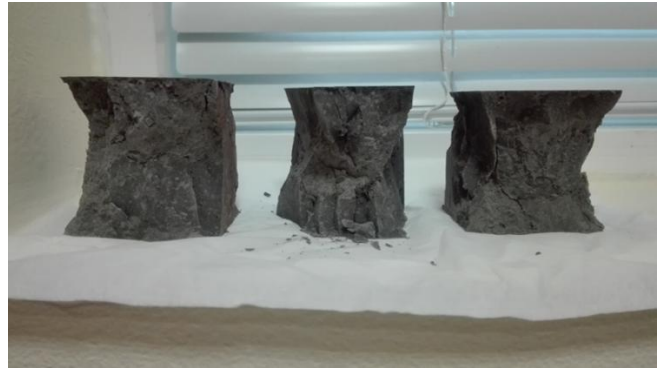

(a) Cube Specimen Ready for Testing           (b) Specimens after Failure

Fig. 2 Specimen testing

### 3. Results and discussions

*3.1 Compressive Strength*

The testing results on the control cement paste samples and the CNT mixed samples are listed in the Table 2 and the Table 3 respectively. The average compressive strength of all the batches of specimens was done on $7^{th}$ and $28^{th}$ day. Control specimens were, however, tested on days 3, 7, 14 and 28. Testing on two days i.e. $7^{th}$ and $28^{th}$ provide reasonable idea about the rate of gain of strength, one being at early age and the other at maturity. Traditionally, the 28 days compressive strength is used in the construction industry for the design of concrete structures. Average compressive strength of control specimens (Table 2) on $7^{th}$ day was about 30 MPa which doubled on the $28^{th}$ day to almost 60.4 MPa. This indicates that the hydration of the cement phases is a slower process. While the cement paste stiffens within few hours, it takes one year for all of the cement to hydrate. However, more than 70% of the cement reacts with water in 28 days[15] therefore a significant development in the compressive strength occurs in this period of time.

**Table 2 Testing results on the control samples**

| Type of Specimen | Specimen ID | Age (Days) | Failure Load (kN) | Compressive Strength, $F_C$ (MPa) | Average $F_C$ (MPa) |
|---|---|---|---|---|---|
| Control | C-a-3 | 3 | 54.3 | 21.72 | 21.69 |
| | C-b-3 | | 57.5 | 23.00 | |
| | C-c-3 | | 50.9 | 20.36 | |
| | C-a-7 | 7 | 76.9 | 30.76 | 29.96 |
| | C-b-7 | | 75.8 | 30.32 | |
| | C-c-7 | | 72 | 28.8 | |
| | C-a-14 | 14 | 95.4 | 38.16 | 37.65 |
| | C-b-14 | | 88.6 | 35.44 | |
| | C-c-14 | | 98.4 | 39.36 | |
| | C-a-28 | 28 | 153 | 61.20 | 60.40 |
| | C-b-28 | | 121.1 | 48.44 | |
| | C-c-28 | | 178.9 | 71.56 | |

The compressive strength of the CNT-mixed cement paste specimens are shown in the Table 3. The specimens with 0.2 wt% pristine CNT exhibited an average compressive strength of 43.4 MPa which increased by 66%, up to a value of 72.2 MPa on 28$^{th}$ day. However, this increase was only 46% in the case of 0.4 wt% pristine CNT specimens. It is worth mentioning that the pristine CNT specimens with 0.2 wt% and 0.4 wt% CNTs had similar strength on 7$^{th}$ day, around 43 MPa. Unexpectedly, 28th day compressive strength of the specimens having 0.4 wt% CNT was found to be 13% less than the specimens with 0.2 wt% CNTs. This may be attributed to excessive replacement of cement in 0.4 wt% CNTs specimens as will be discussed later in detail. The CNTs have no cementitious properties therefore excessive replacement of cement must have led to reduction in compressive strength. In comparison to control specimens, $28^{th}$ day compressive strength of 0.2 wt% regular-CNT specimens increased by 19.5% but for 0.4 wt% specimens this increase was only 5%.

The results are even more interesting for the OH and COOH functionalized CNT mixed cement paste specimens. The 7 day compressive strengths of OH-CNT and COOH-CNT specimens were almost the same as the regular-CNT specimens, i.e. 43 MPa on average but 45% higher than the control specimens. The 28 day compressive strengths of 0.2 wt% OH-CNT and COOH-CNT specimens were 67.7 MPa and 75.1 MPa, an increase of 12% and 24%, respectively. Similar to regular-CNT specimens, reduction in compressive strengths of 0.4 wt% OH-CNT and COOH-CNT specimens was observed. The 0.4 wt% OH-CNT specimen, on day 28, failed at a stress of 53 MPa which is even less than the control specimen by almost 9%. The strength of 0.4 wt% COOH-CNT was no different. It failed at a stress of 57.7 MPa, a value 4.5% less than the control specimen. The absolute maximum compressive strength on day 28 was observed for 0.2 wt% COOH-CNT specimens, which failed at 75.1 MPa, an increase of 24% with respect to the control specimen.

**Table 3 Testing results on the CNT mixed cement paste samples**

| Type of Specimen | Specimen ID | Age (Days) | CNT concentration (wt%) | Failure Load (kN) | Compressive Strength, $F_C$ (Mpa) | Average $F_C$ (Mpa) |
|---|---|---|---|---|---|---|
| Pristine CNT | P-2-a | 7 | 2 | 107.5 | 43 | 43.40 |
| | P-2-b | | | 104.2 | 41.68 | |
| | P-2-c | | | 113.8 | 45.52 | |
| | P-4-a | 7 | 4 | 115.7 | 46.28 | 43.28 |
| | P-4-b | | | 101.5 | 40.6 | |
| | P-4-c | | | 107.4 | 42.96 | |
| | P-2-a | 28 | 2 | 179.8 | 71.92 | 72.21 |
| | P-2-b | | | 170.4 | 68.16 | |
| | P-2-c | | | 191.4 | 76.56 | |
| | P-4-a | 28 | 4 | 163.3 | 65.32 | 63.39 |
| | P-4-b | | | 156.5 | 62.6 | |
| | P-4-c | | | 155.6 | 62.24 | |
| OH-CNT | OH-2-a | 7 | 2 | 117.3 | 46.92 | 45.37 |
| | OH-2-b | | | 112.8 | 45.12 | |
| | OH-2-c | | | 110.2 | 44.08 | |
| | OH-4-a | 7 | 4 | 112.9 | 45.16 | 46.52 |
| | OH-4-b | | | 115.7 | 46.28 | |
|  | OH-4-c |  |  | 120.3 | 48.12 |  |
|  | OH-2-a |  |  | 170.4 | 68.16 |  |
|  | OH-2-b | 28 | 2 | 164.5 | 65.8 | 67.69 |
|  | OH-2-c |  |  | 172.8 | 69.12 |  |
|  | OH-4-a |  |  | 134.6 | 53.84 |  |
|  | OH-4-b | 28 | 4 | 127.7 | 51.08 | 53.03 |
|  | OH-4-c |  |  | 135.4 | 54.16 |  |
| COOH-CNT | COOH-2-a |  |  | 105.3 | 42.12 |  |
|  | COOH-2-b | 7 | 2 | 102.5 | 41 | 42.44 |
|  | COOH-2-c |  |  | 110.5 | 44.2 |  |
|  | COOH-4-a |  |  | 109.4 | 43.76 |  |
|  | COOH-4-b | 7 | 4 | 119.9 | 47.96 | 45.71 |
|  | COOH-4-c |  |  | 113.5 | 45.4 |  |
|  | COOH-2-a |  |  | 191 | 76.4 |  |
|  | COOH-2-b | 28 | 2 | 187.3 | 74.92 | 75.13 |
|  | COOH-2-c |  |  | 185.2 | 74.08 |  |
|  | COOH-4-a |  |  | 157.3 | 62.92 |  |
|  | COOH-4-b | 28 | 4 | 142.6 | 57.04 | 57.73 |
|  | COOH-4-c |  |  | 133.1 | 53.24 |  |

The summary of the compressive strength tests is illustrated in the Fig. 3. The trend clearly shows that although the CNTs doping for three types does enhance the strength of the hydrated cement paste in all the cases however it is true for only certain value (0.2 wt%) of concentration. Increasing the CNTs concentration, indeed has a negative effect on the strength.

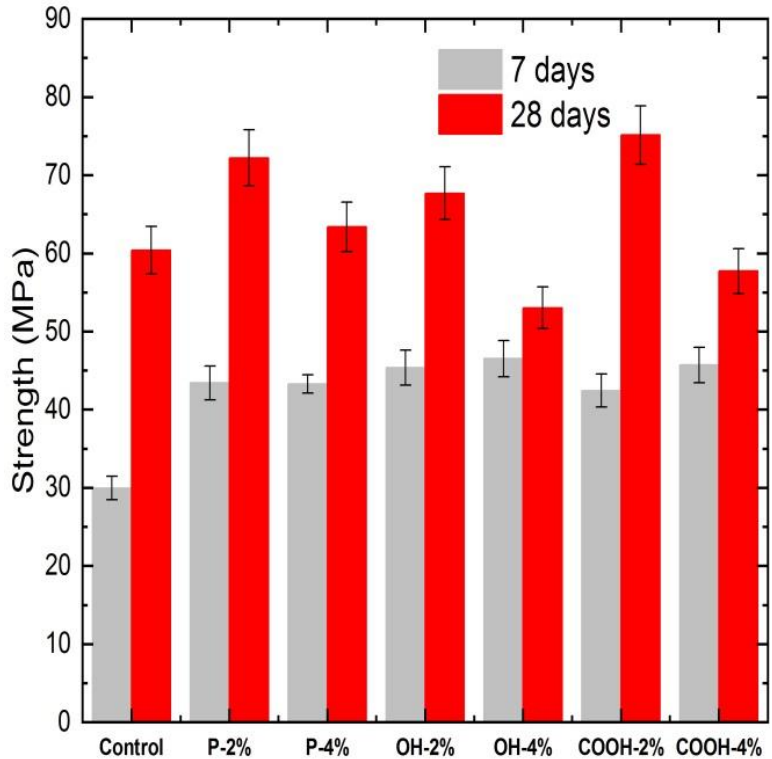

Fig. 3 Comparison of Compressive Strengths for all the Specimens

*3.2 Carbon nanotubes - Scanning Electron Micrographs and Raman Spectra*

The micrographs were obtained from Tescan VEGA XM variable pressure scanning electron microscope. The three types of CNTs are shown in the Fig. 4(a-c). Apparently the CNTs are twisted and kinked in the form of bundles due to strong van der Waal forces. Disentanglement of the CNTs and their uniform dispersion in the cement paste requires a particular attention. In addition, there is a high probability of the presence of the structural and morphological defects in the CNT bundles, also confirmed by Raman spectroscopy.

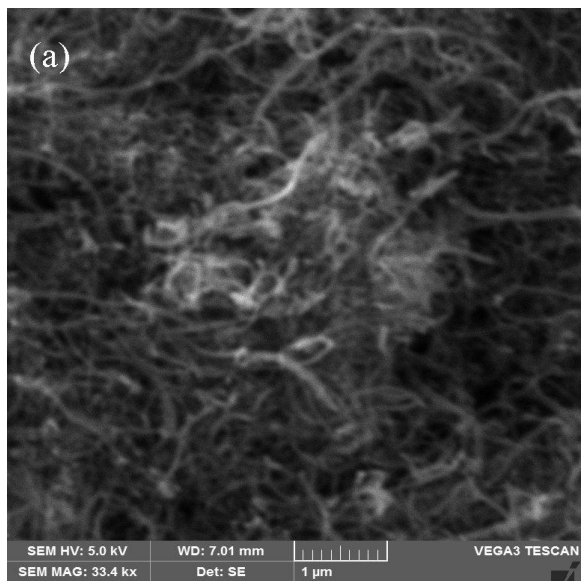
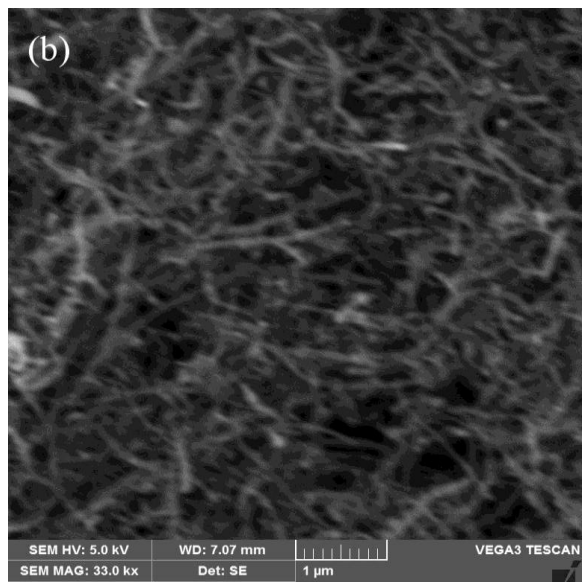
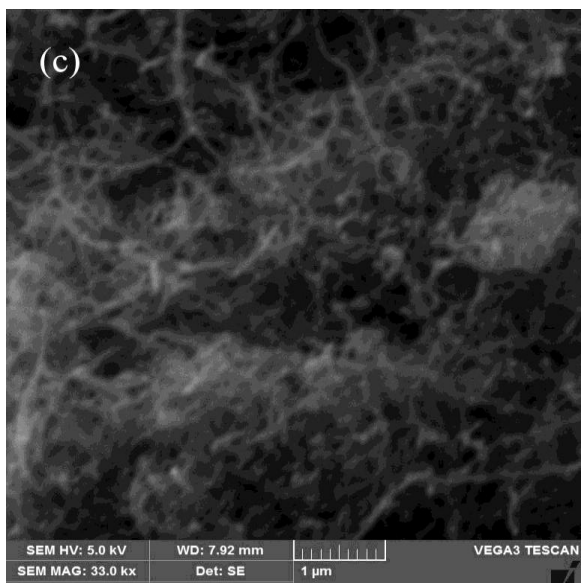

Fig. 4 Scanning electron microscopy of CNTs (a) pristine (b) OH functionalized (c) COOH functionalized

The Raman spectra was collated at different sites of a single bundle and then the average was calculated for all the spectra. The spectra for the three types of CNTs is shown in the Fig. 5. The absence of the radial breathing modes (RBM), the first order Raman scattering process, at lower frequencies (less than 500 cm$^{-1}$) and broader peaks indicate that the CNTs are multilayered. The RBM signals are usually associated with single walled nanotubes (SWNT) of small diameter. Multi-walled nanotubes (MWNT) on the other hand are several SWNTs of variable diameters and chirality wrapped on top of each other seamlessly. The RBM signals therefore are weaker due to incoherent out of phase stretching of bonds.

The most prominent peak at 1582 cm$^{-1}$ is the so called G-band, another first order Raman process which is a finger print feature shared by the graphite, SWNT and MWNTs. However there are subtle difference in the shape of peaks for the each type. The G-band is weakly asymmetric for MWNTs compared to a clear Lorentzian centered at the at a value of Raman shift 1582 cm$^{-1}$ for 2D graphite.

Furthermore, the D and D overtone bands are also visible at 1353 cm$^{-1}$ and the 2708 cm$^{-1}$ respectively. These bands are associated with the defects and the doping respectively and therefore can be used to characterize and monitor the structural changes in the CNTs[16]. For example for the spectra shown in the Fig. 5, the D/G ratio for the COOH and OH functionalized is almost 0.70 whereas for the pristine CNTs, the ratio is approximately 0.50 showing that the order of defects is higher in the functionalized CNTs. These defects play a pivotal role in developing a long range electrostatic and crystal order in the cement paste.

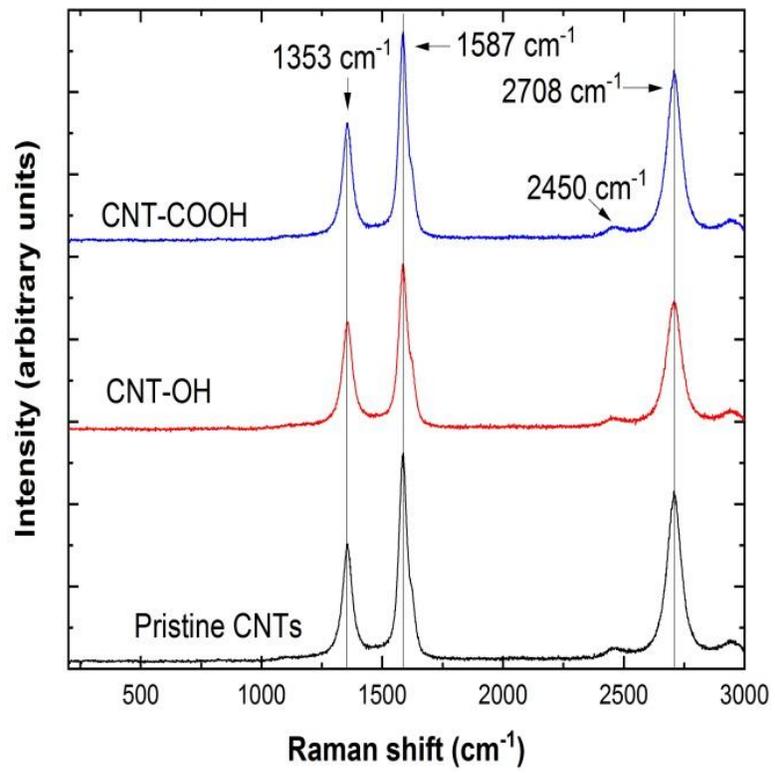

Fig. 5 Raman spectra from carbon nanotubes.

*3.3 Cement, cement paste and cement paste mixed with CNTs – Scanning Electron Micrographs, FTIR spectra and XRD patterns*

Representative SEM micrographs for the control specimens of the cement paste and the cement paste mixed with the OH and COOH functionalized CNTs are shown in the Figs. 6-9. It should be noted that the physical appearance of the control samples was of gray color whereas the CNT doped specimens appeared to be of dark gray color. Particularly the samples with higher concentration of CNTs were clearly distinguishable from the low CNTs concentration cement pastes and control samples. Based on their characteristic physical appearance, therefore, samples could be easily identified. It also goes on to show that CNTs were uniformly dispersed in the cement paste.

The Fig. 6a shows the hydrated segment of a 7 days old control specimen. A deep and wide cleavage is clearly visible in the specimen. As discussed above, a 7 days old sample is not completely hydrated as shown in the Fig. 6b and therefore materials is weak. The Fig. 6d shows a micrograph from the 28 days old control sample where lumps of the hydrated cement form voids and pores despite achieving a reasonable hydration of the cement paste. The hydrated and un-hydrated phases form small clusters therefore the cracks are uneven and in random directions. Fragments of the micro-sized particles are compacted together loosely in the form of lumps. Since there is still 20%-30% un-hydrated cement present in the paste so it is highly likely that the breaking process also involves slipping of the adjacent grains as seen in the Fig. 6d. These features indicate a weak composite structure.

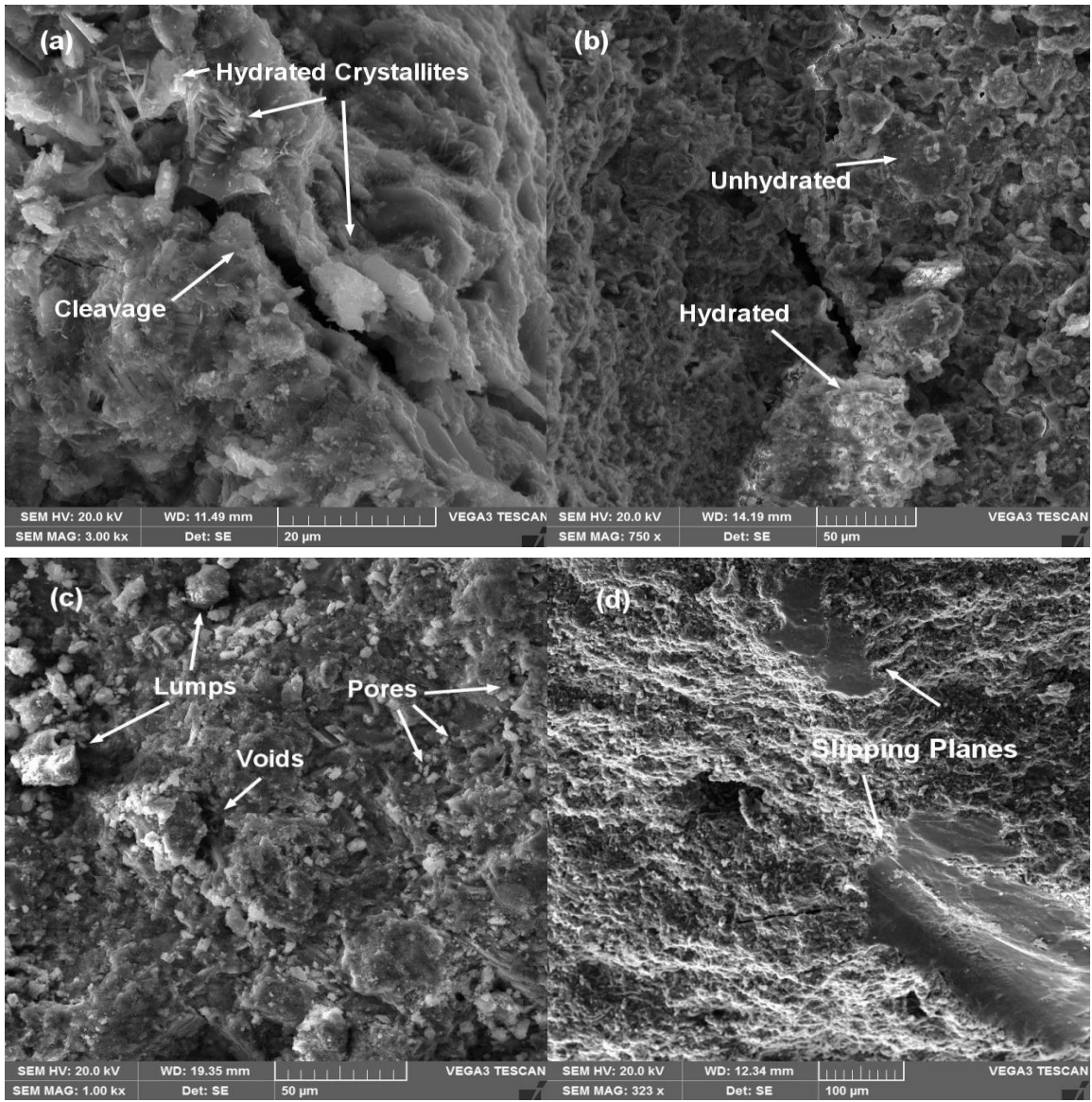

Fig. 6 SEM micrographs for control specimens (a, b) 7 days old and (c, d) are 28 days old.

The cement paste specimen mixed with the pristine CNTs, on the other hand, are remarkably distinct. The 7 and 28 days old specimens prepared with 0.2 wt.% CNTs are shown in the Fig. 7a & 7b. The Fig. 7a clearly indicates the presence of the large number of the small hydrated crystallites for a 7 days old specimen although it is not completely hydrated. Upon achieving a complete hydration, the crystallites grow bigger as shown in the Fig. 7b. The voids and the pores are almost filled due to the formation of the large clusters of the hydrated crystals. The cracks are narrowed down and shallow. The formation of the large hydrated crystals is assisted by the presence of the pristine CNTs which provide a platform for the growth of the crystals.

The specimens prepared with the 0.4 wt% of pristine CNTs offer an interesting perspective though. The hydrated crystallites still grow on the CNT sites as shown in the Fig. 7c & 7d (7 and 28 days old respectively) but the size of the crystallites is not big. The increased density of the CNTs in the cement paste has provided more sites for the crystallites to grow but for the same cement-to-water ratio, therefore the crystallite size is reduced. This trend is shared by all the specimens prepared at the CNTs concentration of 0.4 wt% as will be shown later. Smaller crystallites would naturally leave voids making the structure weaker. This is why the specimens with the higher concentrations of CNTs conceded to a significantly lower force.

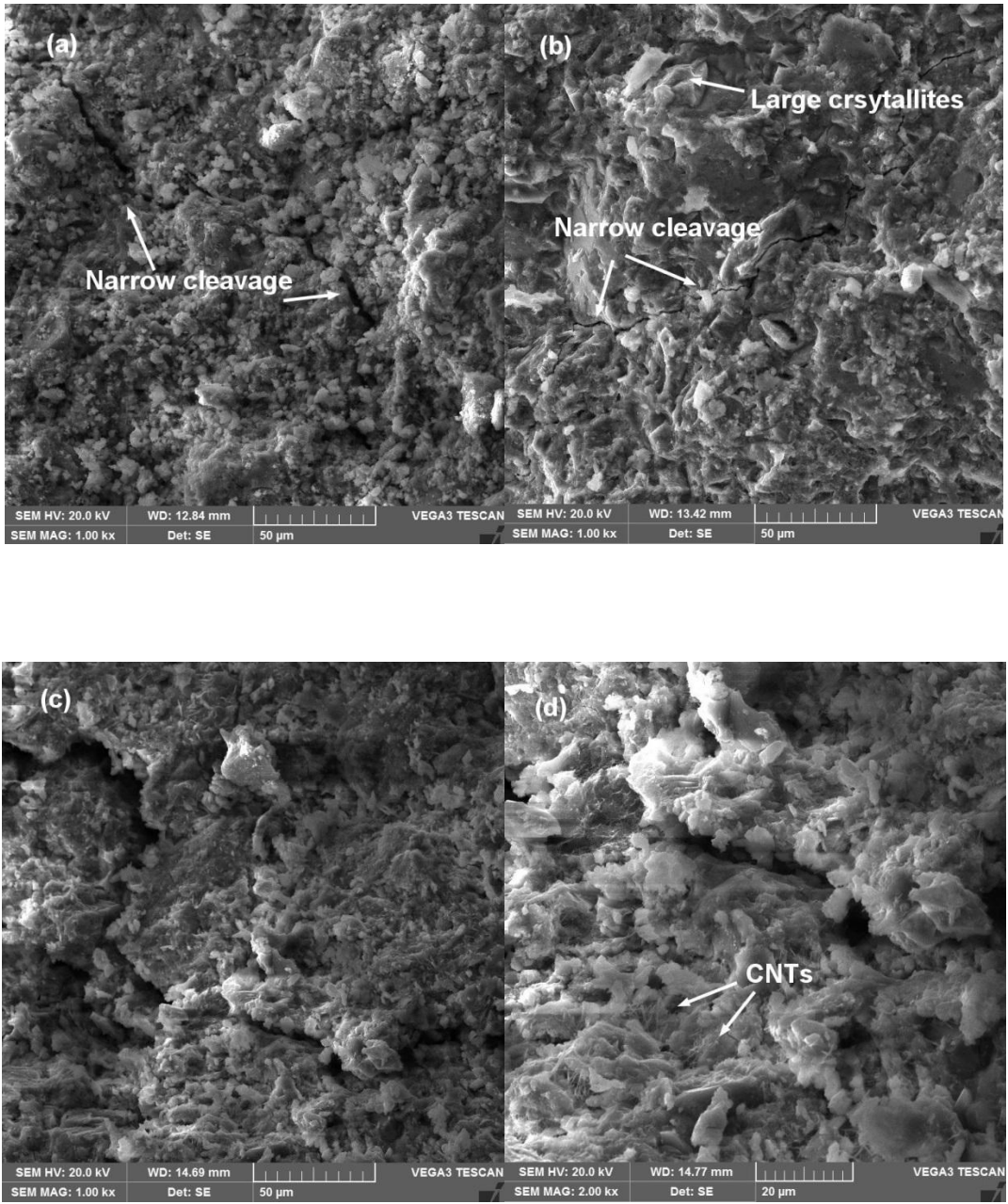

Fig. 7 SEM micrographs for cement paste mixed with the pristine CNTs. Specimens with the 0.2 wt% of CNTs (a) 7 days old, (b) 28 days old. Specimens with 0.4 wt% CNTs (c) 7 days and (d) 28 days old.

The behavior of the concrete specimens mixed with the hydroxyl (OH) functionalized CNTs is very interesting. The 7 and 28 days old specimens of 0.2 wt% CNTs mixed with the cement paste, shown in the Fig. 8a and 8b, were similar to the other specimens (see Fig. 7a and 7b for example) therefore, as expected, their response was the same. However, the cement paste specimens mixed with 0.4 wt% of CNTs exhibited an anomaly. The micrographs of such samples are shown in the Fig. 8c and 8d where small sized crystallites of the hydrated products are visible, however, in contrast to the rest of the specimens, these crystals do not grow further. Strongly bonded OH groups prohibit the further growth of the crystals. The adsorption of the cationic and anionic charges on the hydroxyl functionalized CNTs is not effective resulting into poor exothermic process which inhibits the growth of the crystals. As a matter of principle, this would produce a weaker structure. The yield point for these specimens (Table 3) appears at even lower point on the compressive strength scale making them the weakest structures in our experiments.

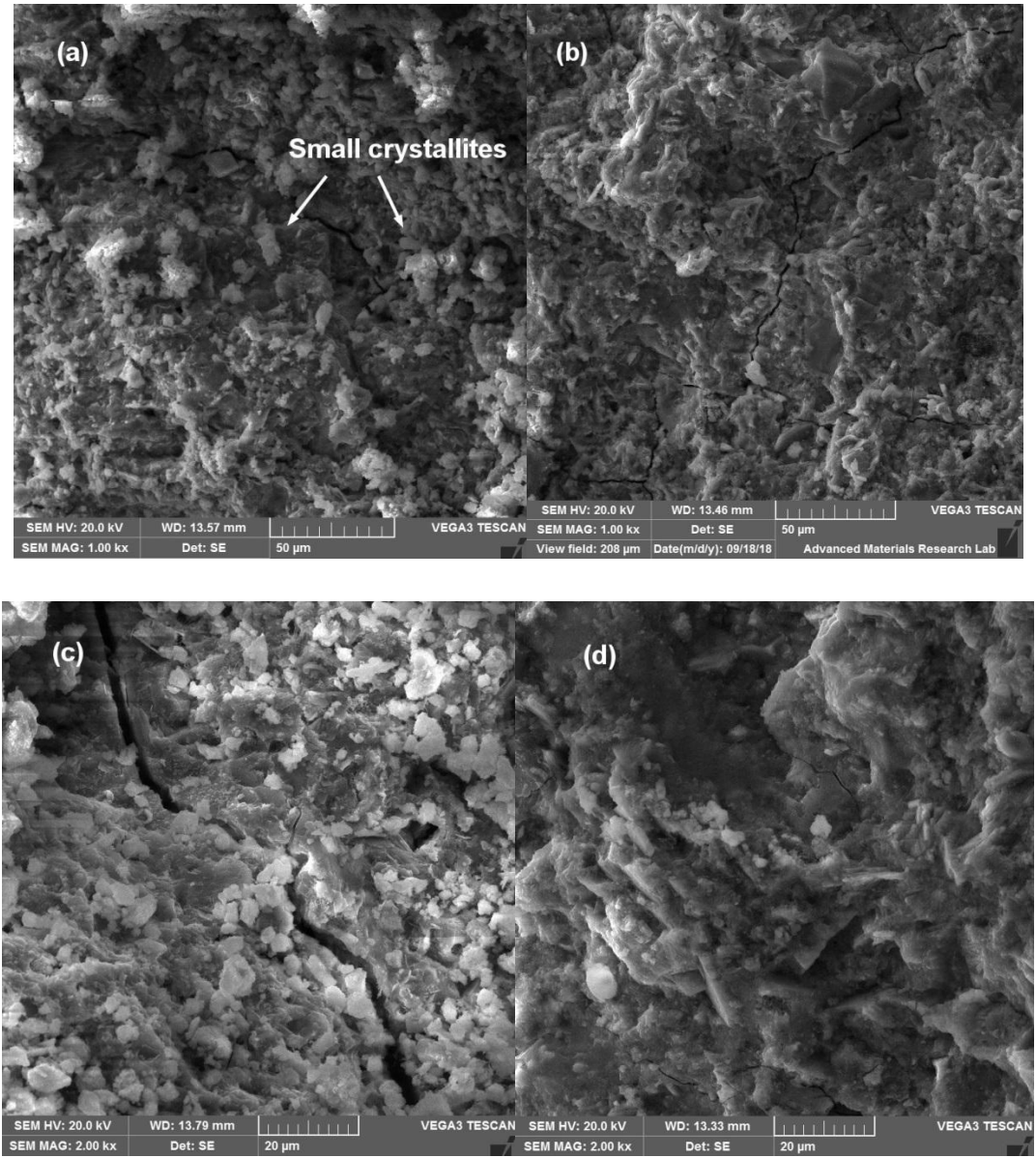

Fig. 8 SEM micrographs for cement paste mixed with the hydroxyl (OH) fucntionalized CNTs. Specimens with the 0.2 wt% of CNTs (a) 7 days old, (b) 28 days old. Specimens with 0.4 wt% CNTs (c) 7 days and (d) 28 days old.

The concrete specimens mixed with the carboxyl (COOH) functionalized CNTs have produced unique results. The addition of the 0.2 wt% of COOH functionalized CNTs increases the compressive strength of the cement paste substantially. These are the strongest specimens in the series of the experiments. The weakly bonded H atoms in the COOH group are easily stripped off leaving the strong negative -COO- charge on the surface of the CNTs. Crystallization of the ionic compounds present in the cement paste is thus facilitated by highly negatively charged CNTs surface providing a long range order to the crystal growth. This is exactly what we see the SEM micrographs in the Fig. 9a and 9b. The size of the hydrated crystallites is large and continuous. In some cases the cleavages appeared to be covered by the crystallites which could be misinterpreted as the bridging between the grains. However, following the trend, these specimens also give up early when the concentrations of COOH-CNTs is increased up to 0.4 wt%. Therefore the bridge formation and pore filling could not be a likely explanation. The Fig. 9c and 9d shows that for a higher COOH-CNT concentrations, the crystallite size is reduced for these specimens as well. As a matter of fact, the breaking point for the specimens mixed with 0.4 wt% of OH-CNTs and COOH-CNTs lie at the same value of average force within 5% experimental uncertainty. Therefore for a concentration of 0.2 wt%, COOH-CNTs mixed specimens are the strongest but as the concentration increases to 0.4 wt%, the strength reduces even below to that of control specimens.

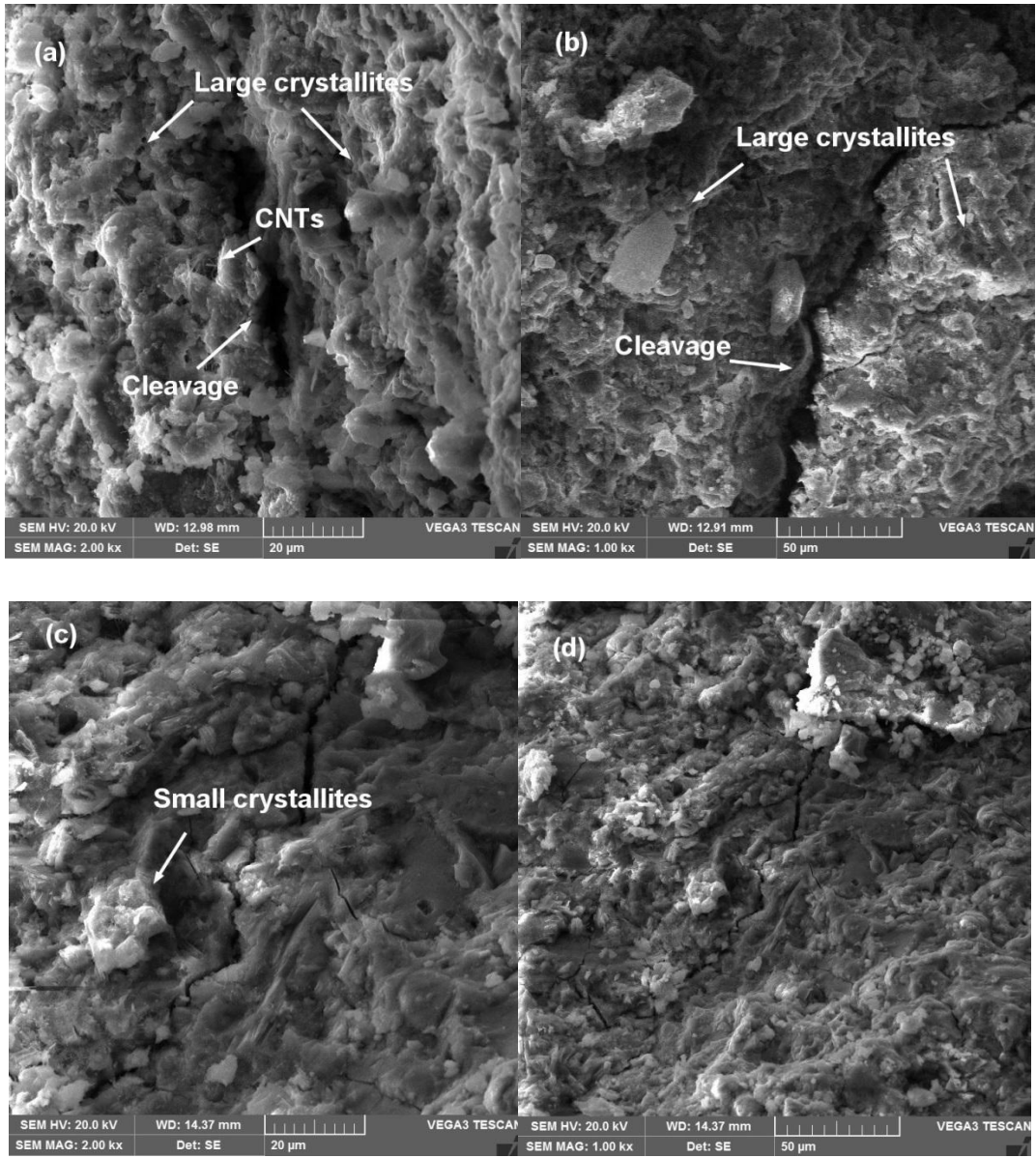

Fig. 9 SEM micrographs for the cement paste mixed with the carboxyl (COOH) functionalized CNTs. Specimens with the 0.2 wt% of CNTs (at) 7 days old, (b) 28 days old. Specimens with 0.4 wt% CNTs (c) 7 days and (d) 28 days old.

The XRD pattern for the cement powder, shown in the Fig. 10, illustrates that the most of the cement is composed of tricalcium silicate ($3CaO \cdot SiO_2$), otherwise known as alite. This is one of the major phase responsible for the strength of the cement on the larger part. Phases of calcium carbonate ($CaCO_3$) and silicon dioxide ($SiO_2$) can also be identified as one of the major, though not very dominating quantitatively. The other unidentified but weaker phases are the oxides of Al, Fe and K etc. The average crystallite size of the powder is almost 40 nm. The random orientation of the crystallite shows that the grains are mostly rough and angular.

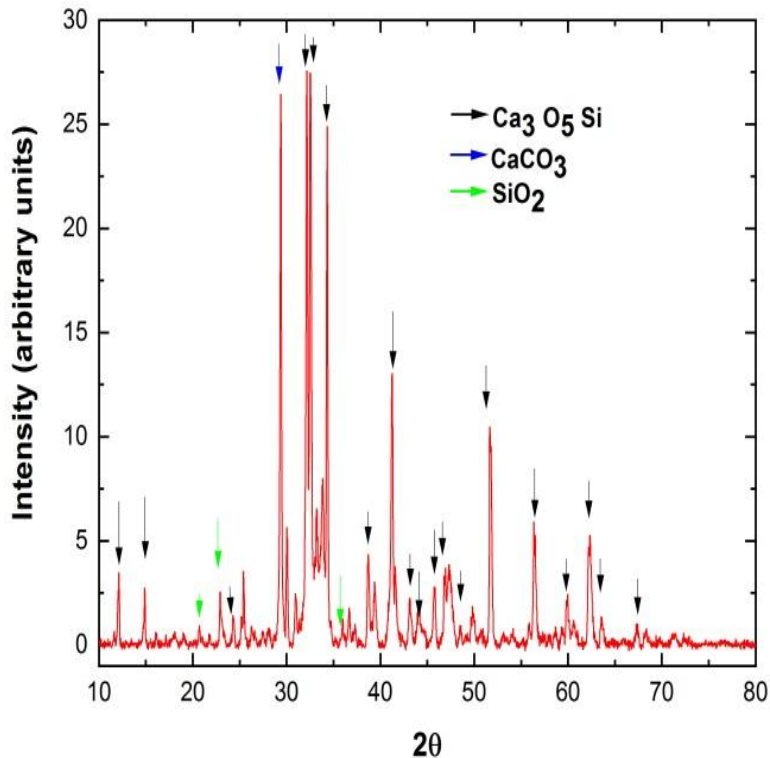

Fig. 10 The XRD pattern for the powder cement

The XRD pattern for the cement paste, and the paste mixed 0.2 wt% and 0.4 wt% of pristine CNTs is shown in the Fig. 11. The specimens mixed with hydroxyl (-OH) and carboxylic (-COOH) carbon nanotubes are not shown here as they were identical to the pristine CNTs. The top panel of the Fig. 11 clearly shows that the powdered cement has reacted with the water to form hydrates. The one of the main hydrate phases formed is the portlandite with a chemical formula of $Ca(OH)_2$. In addition, the other major phases visible are larnite ($Ca_2O_4Si$) and ettringite ($Al_2Ca_6H_{64}O_{50}S_3$) whereas the ferrite and aluminate phases overlap and are indistinguishable at this level of the resolution. Apparently the features of the top panel are shared by the middle and bottom panel both qualitatively and quantitatively, indicating that there has been no chemical reactions between the phases of the cement and the CNTs. The enhanced strength, thus, has not come from any new chemical compound formed within the cement paste.

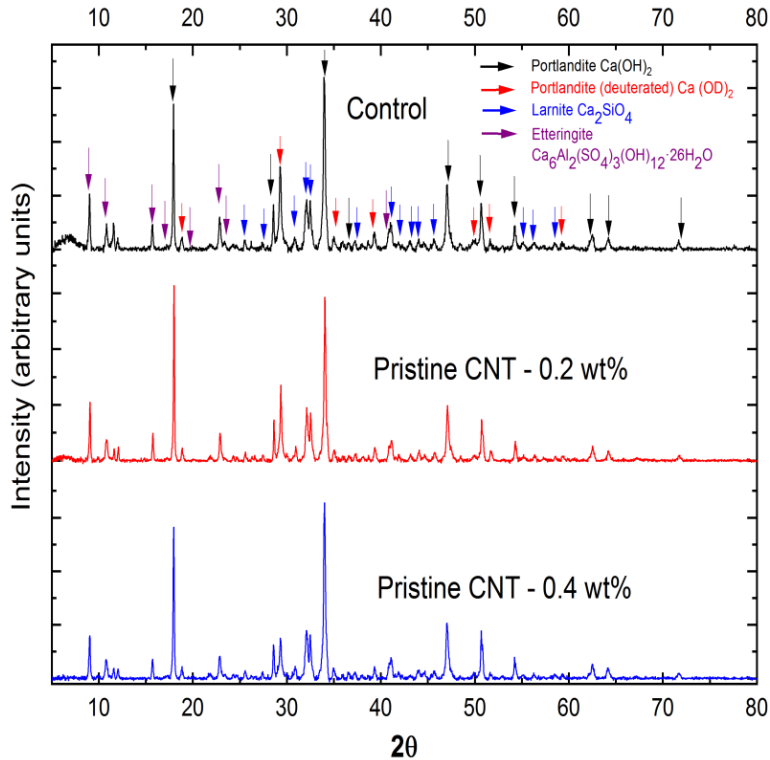

Fig. 11 XRD patterns for the hydrated cement (top panel), cement paste mixed with pristine CNTs at a concentrations of 0.2 wt% (middle panel) and of 0.4 wt% bottom panel.

The above discussion makes it imperative to look for an alternative explanation to understand the role of the CNTs in the strengthening mechanism of the cement paste. The clue appears from the FITR spectra of the cement paste samples as shown in the Fig. 12. In the fingerprint region (below the wave number 1500 cm$^{-1}$), the FTIR signals have a reasonable agreement with the literature[17–19] and have been discussed at length. Therefore no further interpretation of the signals is required in this region. However the diagnostic region (above 1500 cm$^{-1}$) requires some careful attention. Only the stretching vibrations that would produce a change in the dipole moment of a molecule will be registered as IR signals in the spectrum. Therefore an IR signal from a symmetrical molecule would not appear in the spectrum. The signal associated at the frequency of 3418 cm$^{-1}$ in the Fig. 12 is considered to be the contribution of hydrogen bonds, particularly -OH bonds of Ca(OH)$_2$ where -OH ions are symmetrically located around the Ca$^{+2}$ ion (Fig. 13). Yet a broad signal of significant intensity appears in the IR spectra of almost all the specimens. It appears to be particularly stronger for the control specimens and for the specimens prepared with 0.2 wt% of CNTs but very weak for higher CNTs concentrations and is absent from the cement powder specimen for the obvious reasons. This is a clear indication that the Ca(OH)$_2$ is asymmetrical in the cement paste. The symmetry is affected by the presence and concentration of the CNTs. An asymmetric molecule has a non-zero dipole moment and thus have an electrostatic interaction with the molecules and nano-objects adjacent to it.

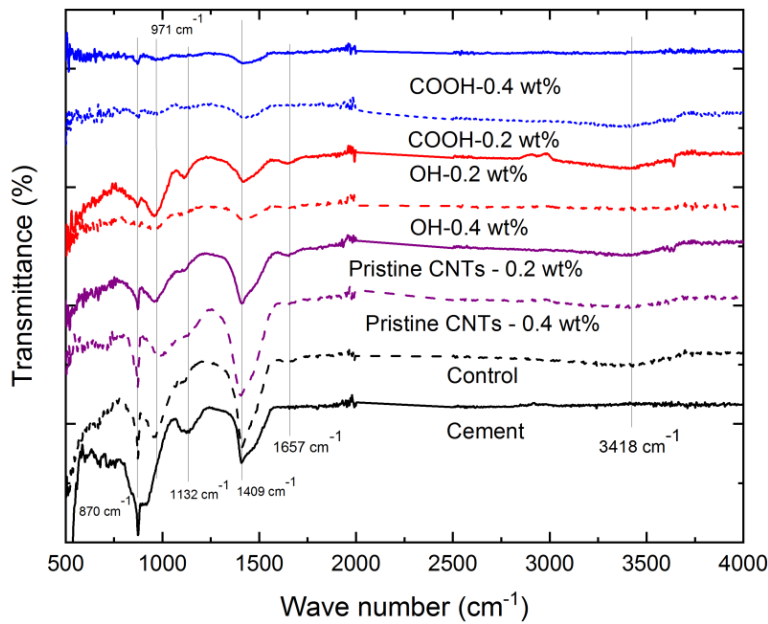

Fig. 12 FTIR spectra for the cement and cement paste specimens mixed with CNTs

## 4. The Theory

The electrostatic interaction between the different compounds and molecules can be mapped through the electrostatic potential. In analogy to a single charge, the electrostatic potential of a molecule sets up an electric field in which it interacts with the other particles/molecules through the Coulomb force. Moreover, since there is a charge accumulation in the ionic as well as covalently bonded molecules, the dipole moment arises which is measured by the magnitude of the charge times the separation between the charges. Higher is the electronegativity of an atom, more it has the ability to keep the electrons closer to itself resulting in electron density build-up at the specific regions around a molecule leaving other regions partially positively charged.

Since the $Ca(OH)_2$ is the major compound found in the cement paste, therefore its interaction with a pristine CNT was modeled. The equilibrium geometry, electrostatic potential and dipole moment were calculated by using density functional theory (DFT) with the functional ωB97X-D and the polarization basis set 6-31G*. The DFT assumes that the exact energy of the electronic states may be described as a function of electron density. The ωB97X-D is a range-separated hybrid (RSH)-generalized gradient approximation (GGA) functional that is not separated into exchange and correlation parts. The functional can be used for the long range non-bonded interactions. The polarization basis set is 6-31G* allows the displacement of electron distributions away from the nuclear positions thus polarizing the elements. This allows us to construct the electrostatic potential surface and dipole moments for the molecules.

In bulk materials systems, where there are large numbers of dipolar ions and molecules interacting with each other, the probability of the random orientation of dipoles is small. The dipoles tend to align themselves and interact with their nearest neighbors in such a way that the overall potential energy of the material system is minimized. It is consistent with the fact that the hydration process is strongly exothermic bringing the hardened material in thermodynamic equilibrium with the environment, and thus increasing its entropy.

The calculated electrostatic potential surface of portlandite, $Ca(OH)_2$, is shown in the Fig. 13. The calcium is doubly positive ion ($Ca^{+2}$) which donates is two electrons to two hydroxide ($OH^-$) ions leaving them negatively charged. Note that the both $OH^-$ ions are on the opposite sides of $Ca^{+2}$ therefore the net dipole moment for the $Ca(OH)_2$ is zero.

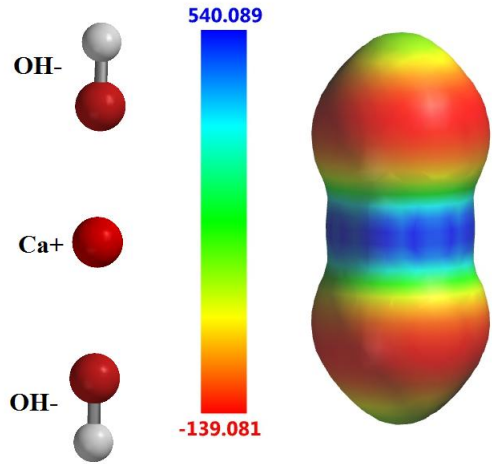

Fig. 13 The arrangement of ions (left) and the electrostatic potential surface of (right) for portlandite

Further, the electrostatic potential surface for a segment of a CNT, shown in the Fig. 14, illustrates the electronic charge density is higher where CNT is terminated. The effective charge on a CNT segment is almost zero and the dipole moment is very weak. However, in an inhomogeneous materials, CNTs are far from their perfect geometry. Instead, they are likely to be under stress from the adjoining crystallites as illustrated in the Fig. 15.

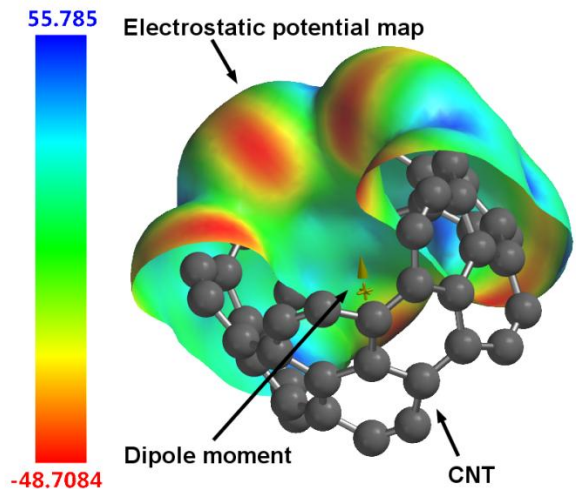

Figure14 The arrangement of the atoms in a CNT and the electrostatic potential surface.

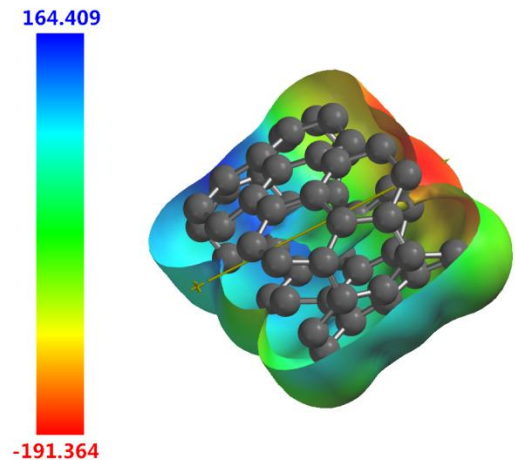

Fig. 15 Model of stressed segment of a CNT and the resulting electrostatic potential surface

Note that in this configuration there is a charge imbalance and the dipole moment is significantly large, nearly 8 D, indicating that the electrostatic force applied by the molecule is strong. The Fig. 16 shows the implications of the electrostatic interaction between a segment of a pristine CNT and a molecule of $Ca(OH)_2$. As discussed above, the $Ca(OH)_2$ and CNT have symmetrical molecular structures however due to Coulombic interaction between the molecules, the molecular symmetry is broken for $Ca(OH)_2$ and the dipole moment of the $Ca(OH)_2$-CNT composite system becomes strong. It is well understood that the ion-dipole interaction is stronger than the dipole-dipole interaction. Therefore the increase in the concentration of CNTs in the cement paste will actually enhance the dipole-dipole interaction between the closest CNT molecules therefore the ion-ion and ion-dipole interaction density will reduce. This is the reason that the increased concentration of CNTs in the cement paste has a negative effect on its. Due to strongly exothermic hydration process and high entropy of the crystallites, the electrostatic interaction is a long range thus increasing the overall strength of the material.

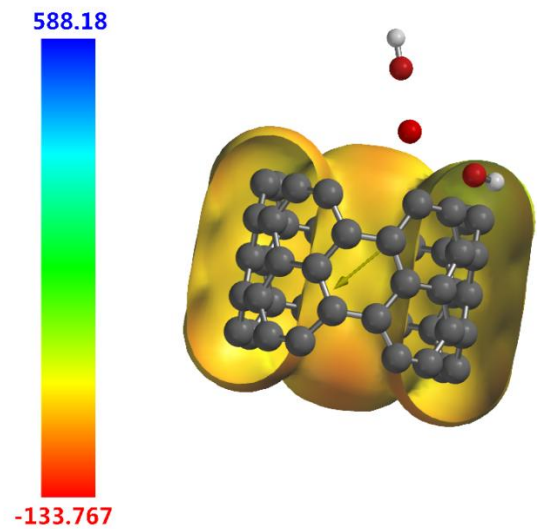

Fig. 16 Interaction of Ca(OH)$_2$-CNT electrostatic interaction.

## 5. Conclusions

The specimens of the cement paste mixed with various proportions and types of CNTs were tested for their compressive strength. The specimens prepared with the 0.2 wt% of CNTs came out to be stronger compared to the ones with no CNTs. However, the specimens became weak by increasing the proportion of CNTs to 0.4 wt%. The SEM micrographs show that the hydrated crystals in the cement paste grow at the CNTs sites. Further, the order of the crystal growth is long range in the specimens with 0.2 wt% CNTs which enhances the strength of the cement paste. On the other hand, increased concentration of CNTs in a specimen provides more sites for the crystal growth inhibiting the development of the long range. Moreover, small crystal growth was observed in the specimens mixed with the hydroxyl (-OH) functionalized CNTs because of strong OH bonds present on the surface of CNTs surface. The concrete samples with the CNTs-COOH functional groups showed growth of large crystallites. Removal of H-bonds from the COOH groups leaves strongly negative surface charge on the CNTs on which ionic crystals develop. The SEM micrographs show no clear evidence of bridging between the adjacent grains in the cement paste specimens. The XRD patterns confirmed that the control and CNTs-mixed specimen have similar chemical composition therefore the source of the enhanced strength must be due to electrostatic interaction between the cement paste and CNTs. The presence of a broad signal in the diagnostic region of FTIR spectrum further confirms the observation that the molecules of hydrated crystals are asymmetric because of the electrostatic interaction with the disordered surfaces of the CNTs. Finally the modelled electrostatic interaction between a hydrated molecule and a CNTs does indeed show a strong ion-dipole interaction. The energy release during the hydration increases the entropy of the composite material system which necessarily reinforces the electrostatic interaction.

doi:10.1016/0008-8846(93)90031-4.